\documentclass[compsoc,conference,10pt]{IEEEtran}
\usepackage{cite}
\usepackage{amsmath,amssymb,amsfonts}
\usepackage{lscape}
\usepackage{algorithmic}
\usepackage{graphicx}
\usepackage{textcomp}
\usepackage{bmpsize}
\usepackage{xcolor}
\usepackage{xurl}
\usepackage{lipsum}
\usepackage{amssymb}
\usepackage[colorlinks=true,urlcolor=black]{hyperref}
\def\BibTeX{{\rm B\kern-.05em{\sc i\kern-.025em b}\kern-.08em
    T\kern-.1667em\lower.7ex\hbox{E}\kern-.125emX}}
\usepackage{makecell}

\usepackage{multicol}
\usepackage{multirow}
\usepackage{array}
\usepackage{stackengine}
\newcolumntype{P}[1]{>{\centering\arraybackslash}p{#1}}
\newcolumntype{M}[1]{>{\centering\arraybackslash}m{#1}}

\usepackage{todonotes}
\usepackage{seqsplit}
\usepackage[T2A,T1]{fontenc}
\usepackage[utf8]{inputenc}     
\usepackage[russian,english]{babel}
\usepackage{color}
\usepackage{ulem}

\begin{document}
\sloppy
\title{Conti Inc.: Understanding the Internal Discussions of a large Ransomware-as-a-Service Operator with Machine Learning}

\author{
\IEEEauthorblockN{Estelle Ruellan}
\IEEEauthorblockA{\textit{University of Montreal} \\
\textit{Montreal} \\
estelle.ruellan@umontreal.ca \\
}
\and
\IEEEauthorblockN{Masarah Paquet-Clouston}
\IEEEauthorblockA{\textit{University of Montreal} \\
\textit{Montreal} \\
m.paquet-clouston@umontreal.ca \\
}
\and
\IEEEauthorblockN{Sebastian Garcia}
\IEEEauthorblockA{\textit{Czech Technical University} \\
\textit{Prague}\\
sebastian.garcia@agents.fel.cvut.cz \\
}

}

\maketitle
\thispagestyle{plain}
\pagestyle{plain}

\thispagestyle{empty}

The full paper is now published in the Crime Science Journal (volume 13, article 16) available at: https://link.springer.com/article/10.1186/s40163-024-00212-y

\begin{abstract}
Ransomware-as-a-service (RaaS) is increasing the scale and complexity of ransomware attacks. Understanding the internal operations behind RaaS has been a challenge due to the illegality of such activities. The recent chat leak of the Conti RaaS operator, one of the most infamous ransomware operators on the international scene, offers a key opportunity to better understand the inner workings of such organizations. This paper analyzes the main topic discussions in the Conti chat leak using machine learning techniques such as Natural Language Processing (NLP) and Latent Dirichlet Allocation (LDA), as well as visualization strategies. Five discussion topics are found: 1) Business, 2) Technical, 3) Internal tasking/Management, 4) Malware, and 5) Customer Service/Problem Solving. Moreover, the distribution of topics among Conti members shows that only 4\% of individuals have specialized discussions while almost all individuals (96\%) are all-rounders, meaning that their discussions revolve around the five topics. The results also indicate that a significant proportion of Conti discussions are non-tech related. This study thus highlights that running such large RaaS operations requires a workforce skilled beyond technical abilities, with individuals involved in various tasks, from management to customer service or problem solving. The discussion topics also show that the organization behind the Conti RaaS operator shares similarities with a large firm. We conclude that, although RaaS represents an example of specialization in the cybercrime industry, only a few members are specialized in one topic, while the rest runs and coordinates the RaaS operation. 
\end{abstract}

\renewcommand\IEEEkeywordsname{Keywords}

\section{Introduction}
In the past ten years, there has been an increase in the scale, complexity, and number of ransomware attacks~\cite{ryan_ransomware_2021}. This was facilitated by the rise of ransomware-as-a-service (RaaS) business models, which provide the infrastructure and technology to conduct ransomware attacks~\cite{salvi_raas_2019, meland_ransomware-as--service_2020, maurya_ransomware_2018, alwashali_survey_2021}. There are well-known RaaS operators, among which Conti (formerly Ryuk) stands up as one of the most active and famous ones~\cite{chainalysis_ransomware_2023}.

Conti has been active since early 2020~\cite{yaaracyberintcom_be_2022}, and its ransomware has targeted high-profile organizations, including government agencies, municipalities, healthcare facilities, law enforcement agencies, 9-1-1 dispatch centers and universities~\cite{noauthor_official_nodate,noauthor_ransomware_2020}. Attacks attributed to Conti are known to demand high ransom payments, generally in Bitcoin, while also threatening to publish the victim's data if the payment is not made~\cite{noauthor_conti_nodate-1}. 

In early 2022, following the Russian invasion of Ukraine, the Conti RaaS operator announced its support to the Russian government. This announcement allegedly led to the leak of hundreds of thousands of their internal Jabber chat logs~\cite{noauthor_vx-underground_nodate}. This leak represents a key opportunity to better understand the inner workings of the Conti RaaS operator. However, given the hundreds of thousands of conversations, a manual analysis represents a time-consuming and monotonous task. 

This study uses machine learning algorithms to uncover insights into the organization of Conti. More precisely, it leverages well-established machine learning methods, including Natural Language Processing (NLP) and Latent Dirichlet Allocation (LDA), coupled with visualization strategies, to uncover the main topic discussions in Conti leaked chats. The results of the analysis uncovered five distinct topics: 1) Business, 2) Technical, 3) Internal tasking/Management, 4) Malware, and 5) Customer Service/Problem-Solving that are distributed across discussions. How these topics are distributed among well-known actors is compared with qualitative analyses conducted by other security researchers. The study's key takeaways are:

\begin{itemize}
  \item The discussion topics uncovered highlight the enterprise-like organization of the Conti RaaS operator. 
  \item A significant proportion of Conti discussions are non-tech related; large RaaS operations require a workforce skilled beyond technical abilities.
   \item  Only 4\% of individuals have specialized discussions, while most individuals (96\%) are all-rounders with diverse discussions.
\end{itemize}

The results of the study corroborate the idea that running such a large RaaS operation translates to developing an enterprise-like structure. The importance of non-tech talk, including business discussions, as well as internal tasking and management discussions, shows that the coordination of large RaaS operations requires a workforce skilled beyond technical abilities. Moreover, only a few individuals need to be really specialized in one area, while the rest coordinate the activities between members and customers. Even for cybercrime organizations, the bigger the organization becomes, the more ``all-rounder'' individuals are required to sustain the economic activities. Finally, this study illustrates how to automatically extract actionable information on the organization of a sophisticated cybercrime organization.

The rest of the paper is organized as follows: Section~\ref{sec:problem} presents a short literature review on ransomware-as-a-service and the Conti group; Section~\ref{sec:methodology} outlines the methods and data; Section~\ref{sec:results} presents the results of the study; Section~\ref{sec:discussion} provides a discussion; Section~\ref{sec:limits} presents the limitations and future research; Section~\ref{sec:conclusion} is the conclusion. 

\section{Background and Context}
\label{sec:problem}
This section starts by presenting the state of research on ransomware and the rise of ransomware-as-a-service. Then, what is known on the Conti organization is presented to provide context to the study's topic. 

\subsection{Ransomware-as-a-Service (RaaS) Business Model}

Ransomware attacks have devastating impacts on enterprises worldwide~\cite{brewer_ransomware_2016}\cite{oosthoek_tale_2022}\cite{kamil_rise_2022}. Such attacks refer to an extorting scheme in which an attacker compromises one or several devices and then locks the device(s) or encrypts the files and asks for money in return for either re-accessing the device(s) and/or obtaining the key that can be used to decrypt the files. Since the first known incidence of ransomware, identified as the AIDS Trojan~\cite{peattie1995approaching}, ransomware attacks have become a central threat to information technologies and the topic of several studies aimed at preventing and detecting it~\cite{kirda_unveil_2017}\cite{kok_ransomware_2019}\cite{richardson_ransomware_2017}\cite{scaife_cryptolock_2016}\cite{song_effective_2016} \cite{lee_ransomware_2018}.

In the past ten years, the threat has evolved. In 2015, ~\cite{kharraz_cutting_2015} analyzed 1,359 samples from 15 ransomware families and found that the number of families with destructive capabilities was small. In the same vein, \cite{gazet_comparative_2010} conducted a comparative analysis of 15 ransomware in 2010 and concluded that ransomware attackers relied rather on small attacks for small ransoms, which led to high amounts due to mass propagation. In this study~\cite{gazet_comparative_2010}, the bulk of ransomware attackers rather followed a low-cost and low-risk business model. Since then, there has been an increase in the scale, complexity, and number of ransomware attacks~\cite{ryan_ransomware_2021}. Indeed, according to recent studies, ransomware attackers are now successful at compromising advanced information systems~\cite{kalaimannan_influences_2017}, and they are better at generating revenue through various extortion schemes~\cite{okane_evolution_2018}. 

Yet, such increase in capacities by ransomware attackers may also be due to the rise of \emph{as-a-service} business models that now characterize the cybercrime industry~\cite{huang_systematically_2018, manky_cybercrime_2013, hyslip_cybercrime-as--service_2020}. Specifically, the ransomware-as-service (RaaS) business model provides the infrastructure and technology to conduct ransomware attacks~\cite{salvi_raas_2019, meland_ransomware-as--service_2020, maurya_ransomware_2018, alwashali_survey_2021}. RaaS clients, known as affiliates, can purchase pre-developed ransomware tools to execute attacks. Usually, affiliates will need to connect to a platform, download the ransomware file, conduct the attack, and manage the victims~\cite{hyslip_cybercrime-as--service_2020}. Also, some RaaS operators provide more support to affiliates, such as negotiating ransoms and providing customer support. In some cases, the affiliate and the operator split the profit generated from the attack~\cite{meland_ransomware-as--service_2020}. In the end, RaaS models reduce the barriers to entry into the market but do not completely remove them as affiliates (those who use the service) still need to have good technical knowledge to purchase the service~\cite{meland_ransomware-as--service_2020}.

Recently, a study by~\cite{chainalysis_ransomware_2023} suggested that a small number of affiliates would be responsible for a large number of attacks and these affiliates would work with many RaaS operators. Such concentration was also observed for RaaS operators as, according to~\cite{chainalysis_ransomware_2023}, there exists a few prolific RaaS operators, including Conti. 

Given the scale and professionalization of these prolific cybercrime operators, their structure may resemble that of an enterprise. Admittedly,
~\cite{lusthaus_industry_2018} interviewed over 200 individuals linked to cybercrime and suggested that some cybercrime organizations may now be organized as firms with offices, floors, and works. Such a corporate-like structure develops where the forces of illegality (as defined by~\cite{reuter_disorganized_1983}), and specifically the risks of arrests, are absent. Without the threat of law enforcement, individuals can openly organize~\cite{lusthaus_offline_2021, lusthaus_industry_2018}. Nevertheless, note that most studies point towards cybercrime organizations being rather small and loosely organized \cite{leukfeldt_criminal_2019} \cite{leukfeldt_cybercrime_2014} \cite{leukfeldt_cybercriminal_2016} \cite{ leukfeldt_examining_2020} \cite{leukfeldt_organised_2017} \cite{ leukfeldt_origin_2017} \cite{leukfeldt_use_2017} \cite{leukfeldt_typology_2017} \cite{lusthaus_industry_2018}. Yet, RaaS providers, and at least Conti, seem to be the exception to the rule. Understanding how these groups operate is key to countering their criminal activities. 

\subsection{The Conti RaaS Operator}

Active since 2020, the Conti RaaS operator successfully ran more than 700 campaigns~\cite{etal_leaks_2022}, generating a revenue, in 2021, of over \$2.7 billion in cryptocurrency~\cite{yaaracyberintcom_be_2022}. To spread the ransomware into their victims' network, Conti was known to leverage phishing campaigns or exploit unpatched software vulnerabilities~\cite{umar_analysis_2021, alzahrani_analysis_2022}. Their phishing campaigns usually contained a \texttt{zip} file or a link luring the victims into downloading a Trojan, which provided a backdoor to deploy their ransomware~\cite{alzahrani_analysis_2022}.  

Following the Russian invasion of Ukraine in February 2022, the Conti RaaS operator announced its support to the Russian government, which allegedly led to the leak of over 160,000 messages from their internal jabber chat logs~\cite{noauthor_vx-underground_nodate}. The person responsible for the leak used a newly created Twitter account under @ContiLeaks~\cite{noauthor_conti_2022-1} to release the files, which also include the source code for the Conti ransomware and other internal project source codes that the Conti organization used to facilitate its operations.

Since then, qualitative analyses of the chat log have been conducted by various security researchers from the private industry~\cite{etal_leaks_2022,noauthor_conti_nodate-1,noauthor_conti_nodate,noauthor_conti_2022,kovacs_conti_2022}. These analyses support the idea that Conti is organized as a firm with physical office buildings, a regular pay schedule and predefined departments such as human resources, finance or  reversing~\cite{etal_leaks_2022,noauthor_conti_nodate-1,noauthor_conti_nodate,noauthor_conti_2022,kovacs_conti_2022}. According to ~\cite{etal_leaks_2022}, Conti's structure followed a classic organizational hierarchy, with team leaders who reported to upper management. The operator had more than 100 people on its payroll, and employees were assigned a specific 5-day workweek~\cite{noauthor_conti_2022}.

Recently, according to~\cite{kovacs_conti_2022}, the Conti organization has shut down the ``Conti brand'', transitioning to a different organizational structure involving multiple subgroups~\cite{kovacs_conti_2022}. Still, the leaked chat log represents a golden opportunity to uncover insights into the organization of the Conti RaaS operator beyond these manual qualitative investigations. 

\section{Methods and Data}
\label{sec:methodology}
This section covers the methods and data used to conduct the analysis and is detailed enough so any researchers who wish to reproduce the analysis on the Conti chat log, but also any other data corpus, can do so easily. The data source, data preprocessing (cleanup), and modeling strategy are presented below. The goal of the analysis was to automatically detect the discussion topics of Conti members. To do so, we used 1) NLP to clean the data, 2) LDA topic modeling to create clusters of groups, and 3) data visualizations to extract meanings from the results.

\subsection{Dataset}
The chat files used for the research were extracted from TheParmak GitHub~\cite{theparmak_conti-leaks-englished_2023}, which was one of the first repositories providing an open source access to the Conti chats translated in English using Google and DeepL.

The available chat logs cover the period from June 21, 2020, to March 2, 2022~\footnote{However, note that there is an absence of data from November 16, 2020, to January 29, 2021. Other sources also show a lack of data over the same period of time~\cite{noauthor_complete_2022, noauthor_vx-underground_nodate}. One possibility is that the user behind the data leak may have wanted to purposely omit this data to avoid incrimination or because it contained sensitive data in some way.}. The data consists of 168,711 chats. These chat logs list the discussions of 346 actors, including members of the organization as well as potential affiliates and customers. The files are in a \texttt{JSON} format, and each log contains the date, the sender, the receiver as well as the actual message. They are structured as follows:  

\begin{itemize}[label={}]
  \item ``ts": ``2021-12-11T08:48:06.821161",
  \item ``from": ``Actor 34@q3mcco35auwcstmt.onion",
  \item ``to":``Actor 77@q3mcco35auwcstmt.onion",
  \item ``body": ``hello"
\end{itemize}

We aggregated all chats sent per actor. Table~\ref{tab:chat-corpus} shows a summary of the aggregated chats per actor after the processing. Such dataset is referred to below as the \emph{corpus}. 

\begin{table}[ht]
    \centering
    \caption{Structure of the data after aggregation per actor}
    \label{tab:chat-corpus}
    \begin{tabular}{|c|c|p{4cm}|}
        \hline
        \textbf{Index} & \textbf{Actor} & \textbf{Chat Corpus} \\
        \hline
        0 & Actor 77 & .in place?..in place? .in place? in place?dro [...] \\
        \hline
        1 & Actor 3 & , yes in part).on the road, unstable online.ca [...] \\
        \hline
        2 & Actor 55 & be on line .. unresolved external symbol \emph{CLSID} [...] \\
        \hline
        [...]   & [...]     & [...] \\
        \hline
        344 & Actor 170 & ku \\
        \hline
        345 & Actor 174 & .I don't see any new coders online with me.in [...] \\
        \hline
    \end{tabular}
\end{table}

When chats were posted as a general message in a channel containing several members, they were appearing more than once in an actor’s corpus. For example, if Actor A posted ``hello guys'' in a channel, it would appear X number of times in the actor’s corpus, with X being the number of people in the channel, even though Actor A posted this message only once, as illustrated in Table~\ref{tab:chat-log}. 

\begin{table}[ht]
\centering
\caption{Chat Logs}
\label{tab:chat-log}
    \begin{tabular}{|c|p{2cm}|p{4cm}|}
        \hline
        \textbf{From} & \textbf{To} & \textbf{Chat} \\
        \hline
        A & D & Hello guys \\
        \hline
        A & C & Hello guys \\
        \hline
        A & B & Hello guys \\
        \hline
        [...] & [...] & [...] \\
        \hline
    \end{tabular}
\end{table}

Such repetitive chats were problematic for the model developed below for two reasons: 1) they distorted what an actor ``really'' posted; the actor’s corpus would no longer be accurately representative of an actor’s activity, and 2) they impaired the process of topic creation as a topic is a set of words that are often seen together throughout documents. They were thus removed. Each actor's corpus was then cleaned using Natural Language Processing (NLP), as explained below. 

\subsubsection{Natural Language Processing (NLP)}
To clean the chats, we used Natural Language Processing (NLP). NLP is a subfield of artificial intelligence that focuses on allowing a machine to understand natural language, that is, human language~\cite{chowdhary_natural_2020, raina_natural_2022}. Basically, NLP teaches a machine to learn, understand, and derive meaning from a language. Natural language processing uses various algorithms to learn and follow grammatical rules, which are then used to derive meaning out of words and sentences~\cite{chowdhary_natural_2020, raina_natural_2022}. Some of the most commonly used algorithms are stemming (reducing words to their lexical root), lemmatization (converting a word into its canonical form), and tokenization (dividing the text into meaningful pieces). NLP is used in a myriad of diversified fields such as biology~\cite{ofer_language_2021}, translation~\cite{zong_application_2018}, business intelligence~\cite{vashisht_integrating_2020}, psychology~\cite{henning_machine_nodate} to name a few. 
  
Using NLP algorithms, we were able to clean the chat logs, keeping only relevant words, such as ``hack'', ``pay" or ``malware". To do so, we first used \emph{normalization}, which changed all words to lowercase. Second, we removed all irrelevant material from the text, like stop words, punctuation, and HTML links. Stop words are commonly used words that are not essential to the context or meaning of the sentence: ``I", ``is", ``the", ``you". Third, we tokenized the text, which consisted in dividing the text into meaningful pieces or elements for the algorithm. The message ``I like blue birds" then became ``[like; blue; birds]". Fourth, we lemmatized the text, which is the process of converting a word into its ``canonical form". In other words, ``codes" became ``code" and ``talked" became ``talk''. Thus, words in the third person were changed to the first person, and verbs in past and future tenses were put in the present tense.

This process allowed us to identify some actors who stood out for their small corpus compared to others. Some had 4,000 and plus words, whereas others only had ten relevant words or even two after the data processing. For the algorithm (presented below) to process the meaning of discussions, an actor has to have a substantial amount of chats. Hence, we removed actors whose corpus contained fewer than 100 words, reducing the number of actors for the analysis to 137, each having a corpus of at least 100 relevant words. Descriptive statistics on the final sample of 137 actors are presented in the table below.

\begin{table}[htbp]
    \centering
    
    \caption{Post Processing Descriptive Statistic}
    \begin{tabular}{|c|p{2.5cm}|p{2.5cm}|}
    \hline
        Statistic & Nb of messages sent & Nb of relevant words after processing \\
        \hline
        count   & 137       & 137 \\
        \hline
        mean    & 419.91    & 2940.62 \\
        \hline
        std     & 1067.66   & 6906.91\\
        \hline
        min     & 5         & 101\\
        \hline
        25\%    & 55        & 295\\
        \hline
        50\%    & 132       & 841\\
        \hline
        75\%    & 361       & 2822\\
        \hline
        max     & 9917      & 55817\\
        \hline
    \end{tabular}
\label{tab:descriptive_stats}
\end{table}

\subsection{Latent Dirichlet Allocation (LDA)}

To find the discussion topics of Conti members\footnote{For the purpose of this study, we consider Conti members anyone who participated in the chats}, we computed Latent Dirichlet Allocation (LDA) topic models based on actors' corpus. LDA is a topic modeling method based on a generative probabilistic model for text corpora. It is widely applied with NLP to uncover topics from unordered corpora of documents ~\cite{blei_latent_nodate}. The basic idea behind LDA is that each document is represented as a finite mixture of latent topics, and each topic is characterized by its own distribution over words. So the LDA extracts the latent topics from a corpus of documents and simultaneously assigns a probabilistic mixture of these topics to each document. Thus the topic probabilities provide an explicit representation of a document. Topic models are applied in various fields, including political science~\cite{zhou_comparative_2019}, medicine~\cite{wu_ranking_2011} and cybersecurity~\cite{kolini_clustering_nodate}. 

The LDA model was implemented using mallet 2.0.8~\cite{noauthor_downloading_nodate} and the gensim wrapper~\cite{noauthor_gensim_nodate}. To find the best model, we developed a strategy that combined both the traditional coherence score along heuristic interpretations of the main topics discussed in each cluster. The coherence score helped distinguish topics that were semantically interpretable topics from topics that were simple artifacts of statistical inference. Such score ranges from zero to one, and the higher the score, the better the model should be. 

The clusters found were evaluated through visualizations created with WordClouds (to visualize the most important words) and semantic space using pyLDAvis ~\cite{noauthor_pyldavis_nodate}. For the latter, the clusters were plotted onto a semantic space where two words in the same lexical field or synonyms were correlated and thus ``close'' to each other in the space. The larger the topic cluster, the more conversations actors had about that topic. The more the clusters (and thus words) were far apart the more these clusters had their wn vocabularies. Overlapping clusters had similar vocabularies. This way, a model with no overlapping was considered good. The best model selected had the highest coherence score and the best visual representation, with far-apart clusters. 

\begin{figure}[ht]
    \centering
    \includegraphics[width=0.5\textwidth]{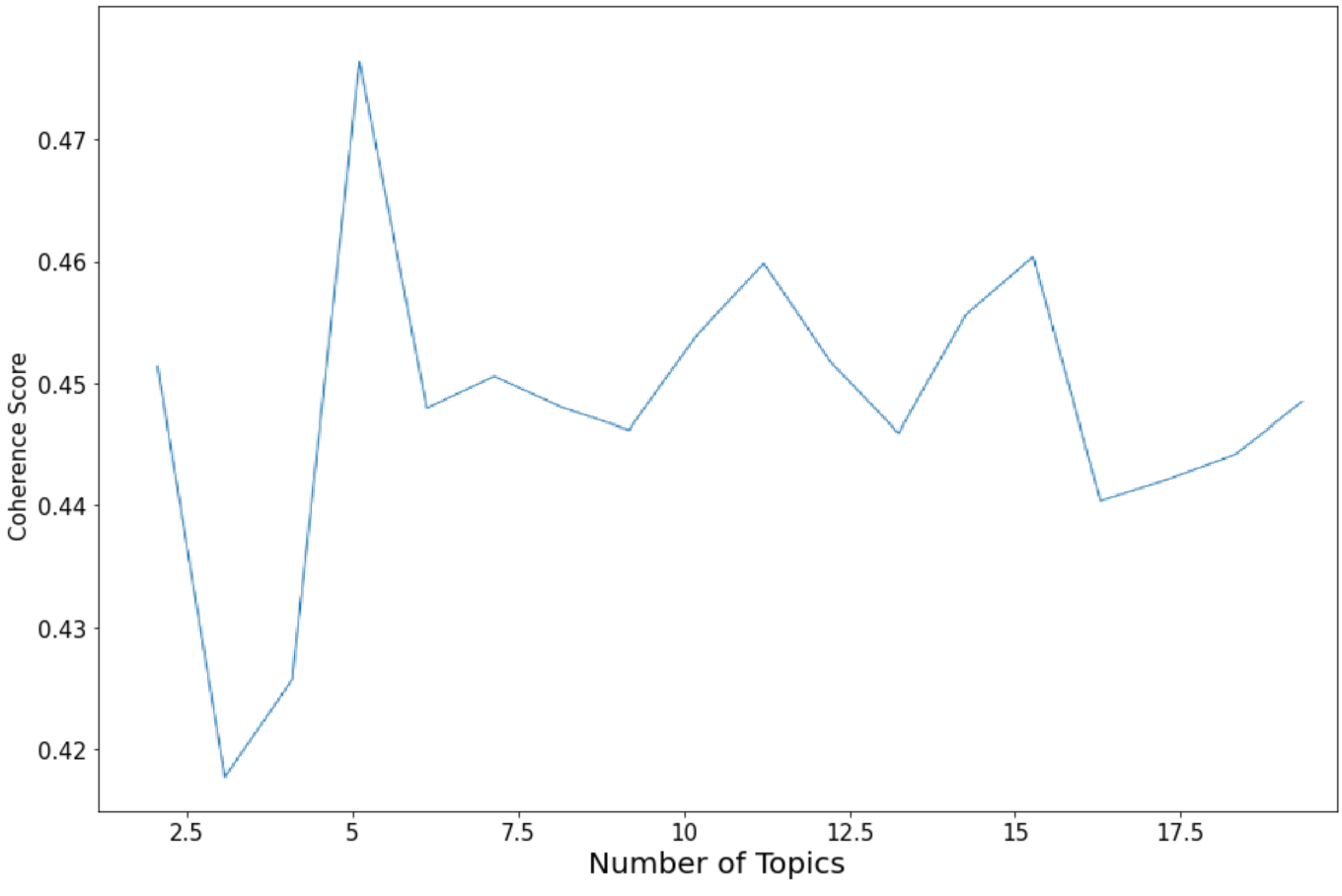}
    \caption{Coherence score per number of topics k}
    \label{fig:coherence_score}
\end{figure}

After training various models with a different number of topics (\texttt{k}), investigating the coherence scores, as shown in Figure~\ref{fig:coherence_score} and inspecting the resulting clusters (with WordClouds and semantic space representation of the topics), the most promising model was the one with \texttt{k=5} topics. 

\subsubsection{Topic distribution:}
    The five topics span across each actors' corpus with different weights as each actor can be represented as a mixture of determined topics: topic 1 may represent 100\% of actor A's corpus, while 60\% of Actor B's corpus. For example, Table~\ref{tab:LDA_representation} shows how the five topics are distributed in Actor 112's corpus and Actor 83's corpus. In this example, Actor 112's discussions revolve clearly around topic 1 whereas Actor 83's discussions revolve around the five topics.

\begin{table}[htbp]
    \centering
    \caption{LDA Representation explained}
    \begin{tabular}{|p{1.25cm}|p{1cm}|p{1cm}|p{1cm}|p{1cm}|p{1cm}|p{1cm}|p{1cm}|p{1cm}|}
        \hline
        Actor & Topic 1& Topic 2& Topic 3& Topic 4 & Topic 5 \\ \hline
        Actor 112 & 98.5\% & 0.3\% & 0.4\% & 0.5\% & 0.3\% \\ \hline
        Actor 83 & 18.6\% & 14.9\% & 23.1\% & 23.9\% & 19.5\% \\ \hline
        \end{tabular}
\label{tab:LDA_representation}
\end{table}

\subsubsection{Topic interpretation:}
The LDA model gives topics that are composed of a word list, often appearing together within chats. It is the researcher’s role to make sense of these topics by giving them a theme or a name based on what they are made of. To do so, we went over the words in the five clusters, interpreting their meaning. We also took the main actors in each cluster (those whose corpus was mainly related to a topic) and read their discussions to have contextual information around the words. The interpretation of the topics is presented below, along with how the topics are distributed among actors. 

\subsubsection{Comparing the study results:}
Finally, to compare the results of the study, we went through summaries of qualitative analyses conducted by security researchers. We found four relevant blog articles by CheckPoint Research~\cite{etal_leaks_2022}, KrebsonSecurity ~\cite{noauthor_conti_2022}, Cyberint~\cite{yaaracyberintcom_be_2022} and Trellix~\cite{noauthor_conti_nodate} that conducted a qualitative analysis on the Conti chat logs to paint a picture of the organization. Each blog article attempts to uncover the roles and importance of each member, providing a description of a few actors identified as key. From these articles, we extracted the role attributed to those well-known actors and compared them with the topic distribution found in this study.  

\subsection{Ethical Considerations}
The study has been approved by the ethics committee at the University of Montreal (project N.2023-4659) under minimal risks. The study required asking for a waiver of consent in line with Article 5.5A of the Canadian Tri-Council Policy Statement on Research Ethics. To ensure participants’ confidentiality and privacy, the real pseudonyms of the actors are not displayed throughout the text.

\section{Results}
\label{sec:results}

The best model included five topics that encompassed actors' discussions. The interpretation of the topics is presented below, followed by how they are distributed among actors' corpus. We then compare the results of this study with previous qualitative research conducted on the role of these actors. 

\subsection{From Business to Tech Topics}

The five topics that span actors' corpus are: 1) Business, 2) Technical, 3) Internal tasking/Management, 4) Malware, and 5) Customer Service/Problem Solving. Each topic is accompanied by an excerpt of a discussion from an actor's corpus whose main topic is the one being presented. \footnote{These excerpts are for illustrative purposes only and do not reflect the format of the actor's corpus provided to the algorithm nor the full range of discussions found within the actor's corpus.}

\textbf{Business topic}: The first topic encompassed discussions regarding planification and internal tasking within a project. Actor 118, Actor 112 and Actor 23 were actors often quoted within chats to repeat what was said or ordered. The topic included words like \emph{build}, \emph{office}, \emph{task}, and \emph{report}, referring to some sort of task management. Words like \emph{system}, \emph{hacker}, \emph{coder}, and \emph{software}, were also included, referring to employees and their work tools. Actors getting the first topic as their dominant topic could be seen as ``higher-ups'' or participating in the management activities of the Conti organization.

Here is an excerpt of a discussion from actor 118, whose main topic is Business: ``\textit{This is an important task, then let's build a system for it [...]. I suggest that you allocate people and build a system that will analyze and report information from these office-based documents, [...] prepare reports by sector, the main department will prepare attacks [...].}"


\textbf{Technical topic}: The second topic revolved around technical talks and developing technical projects. The vocabulary of this topic was very much focused on computer science, including words like \emph{version}, \emph{command}, \emph{module}, \emph{program}, \emph{function}, \emph{system}, \emph{window}. Some other words were even more specific and denoted an attack vector or part of it: \emph{script}, \emph{loader}, \emph{backdoor} and \emph{.exe}. Actors having a tendency towards this topic could be taking part in delivering attacks. 
Here are excerpts from actors 86 and 54's corpus whose main topic is the Technical topic:``\textit{When an error occurs during process hollowing creation, do you send an error code to the server? [...]}" and ``\textit{I tried to shift the .exe file image in the process address space (i.e. to modify the process hollowing) and to write it to an arbitrary address, but this didn't work.}"

\textbf{Internal tasking/Management topic}: The third topic was the only one without any computer science or technical words in it. The core of this topic was about human resources, management, and salaries. The topic included words like \emph{salary}, \emph{people}, \emph{money}, \emph{email}, \emph{network}, \emph{talk}, \emph{team}, \emph{buy}, \emph{month}, \emph{salary}, \emph{touch}, \emph{company}, \emph{blog} and \emph{offer}. The words \emph{onion} and \emph{protonmail\_com} were also there, which are both domains used to communicate or add actors to different channels. Actors holding a high percentage of correspondence to this topic may have been involved in human resources, internal tasking and management tasks.

Here is an example of a discussion from actor 124's corpus, whose main topic is Internal tasking/Management: ``\textit{I'll help you when you get your salary. Add to your contacts Actor 101, this is your team leader. [...] salary pay 2 times a month to your bank card. [...] workday 10-11 to 7:20 p.m, but it's best to discuss this with your supervisor [...].}"

\textbf{Malware topics}: The fourth topic was directed toward one type of attack vector:  malware and/or ransomware. Many of the words that made up this topic alluded to the injection or implementation of the malware as well as stratagems to avoid detection: \emph{DLL} (refers to DLL hijacking), \emph{detect}, \emph{crypto}, \emph{crypt}, \emph{loader} and \emph{pour} (term used as a synonym of launch/inject). An actor having the fourth topic as its main topic was likely taking part in the conception of malware as an attack vector.

Here is an excerpt of a discussion from actors 11 and 85 whose main topic is Malware: ``\textit{As long as it is through rundll32 and dll pathmake [...] with pdf icon. [...] I've run a new version of loader [...]}" and ``\textit{Don't crypt [encrypt files] if you're going to, I'll be pouring in new files soon.}"

\textbf{Customer Service/Problem-solving topics}: The fifth and last topic appeared to be a bit blurrier, including two subtopics. The first revolved around customer service with words like \emph{order}, \emph{payment}, \emph{client} and \emph{receive}. The second subtopic related to what seemed to be attack assistance or problem-solving, with words like \emph{log} (i.e., record of the events), \emph{error}, \emph{module}, \emph{proxy} and \emph{IP}. Actors with this topic as their main topic would represent actors who solved problems while also dealing with clients. 

Here is a quote from actor's 36 corpus whose main topic is Customer Service/Problem-solving: ``\textit{if a lib [library] crashes, it means the client [affiliates] isn't sending what the lib is expecting [...] so the http parser crashes. You should give specifications to those who write to clients, what this lib can and cannot do. This is an industrial solution ... and a lot of people use it.}''.

\subsection{Multifaceted Discussions of Conti Actors}
 
\begin{figure*}[ht]
  \centering
  \includegraphics[width=\textwidth]{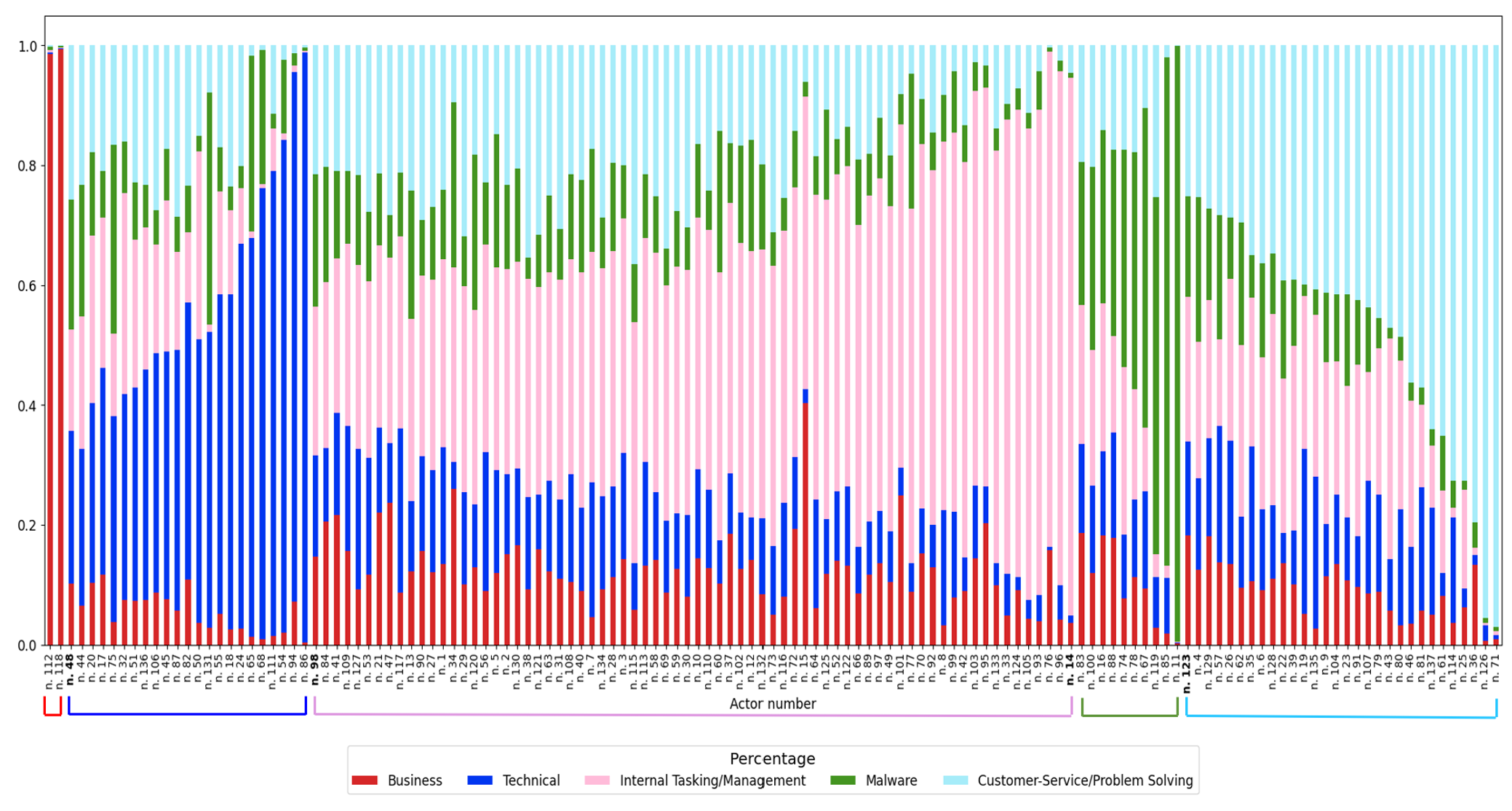}
  \caption{Topic Distribution per Actor}
  \label{fig:stacked}
\end{figure*}

Figure~\ref{fig:stacked} displays the distribution of topics for each actor through a stacked bar graph. The colored brackets grossly emphasize where the prevalence of a topic is high across the actors' corpus. The figure shows that only a small number of actors (including Actor 112, Actor 118, Actor 86, Actor 94, Actor 11, Actor 85, Actor 126, and Actor 71) have discussions that centered around a single topic. For these actors, their stacked bar is almost monochrome, meaning that their discussions were almost entirely focused on a single topic. Quite the opposite, the rest of the studied actors' stacked bar is a mixture of multiple topics, illustrating the diverse and all-rounder discussions that most actors had. 

Figure~\ref{fig:stacked} also shows that the Business [red] and the Malware [green] topics are the rarest ones in members' discussions. Moreover, the number of actors' whose corpus specializes in one of these two topics is small, including Actor 112 and Actor 118 for the Business topic, as well as Actor 11 and Actor 85 for Malware topic. 

In the same fashion, the Customer Service/Problem Solving [light blue] and Technical [dark blue] topics are spread among actors, with a few of them having their discussions centered specifically on one of these two topics. 

On the other hand, the Internal tasking/Management topic [pink] is widely spread among actors. Actually, such topic is present in almost every actor corpus and monopolizes a moderate to high part of actors' discussions. Such topic is not technical (like the Business topic); it included discussions on human resources, management, and salaries. Such result illustrates the intensive non-technical aspect of RaaS operations, which seemed to monopolize time and effort for a large proportion of Conti actors. 

Finally, out of 137 actors, six had specialized discussions with 95\% of their discussions revolving around a single topic. Table~\ref{tab:special} shows the six actors and the topic they specialized in. In short, the discussion of Actor 118 and Actor 112 were mainly about Business, Actor 11 focused on Malware, Actor 86 on Technical and Actor 126 and Actor 71 on Customer Service/Problem Solving.  

\begin{table}[ht]
    \centering
    \caption{Specialized Actors with Percentage of Dominant Topic in their Corpus}
    \label{tab:special}
    \begin{tabular}{|p{2cm}|p{2cm}|p{2.5cm}|}
    \hline
        Actor       & Dominant Topic    & \% of corpus related to dominant topic\\ \hline
        Actor 118   & Business          & 99.4\%\\ \hline
        Actor 11    & Malware           & 99.3\%\\ \hline
        Actor 86    & Technical         & 98.6\% \\ \hline
        Actor 112   & Business          & 98.5\% \\ \hline
        Actor 71    & Customer Service/Problem Solving  & 97.1\%\\\hline
        Actor 126   & Customer Service/Problem Solving  & 95.6\%\\ \hline
        \end{tabular}
\end{table}

All in all, this means that only 4.38\% of the studied actors were specialized in a single topic, whereas 95.62\% were all-rounders, with a corpus of discussion revolving around the five topics.

\subsection{Topic Distribution of Well-Known Actors}
This section compares the results obtained using machine learning to external sources' results obtained by humans reading the chat logs to assess if our results are coherent. This comparison also serves to evaluate the coherence of the machine learning model's output when compared with human judgment.

To compare the results of this study, we went through previously published blogs in which the Conti chats were analyzed qualitatively and extracted the role of well-known actors according to sources. We present in Table~\ref{tab:existing_analysis} the role assigned to well-known actors by external researchers and their distribution of topics based on the results of this analysis. To facilitate the analysis, we focus on their dominant topics, meaning the topic with the highest percentage in the actor's corpus. 

As shown in Table~\ref{tab:existing_analysis}, the two actors with their dominant topic being Business are Actor 112 and Actor 118. They were both interpreted as being the organization bosses in other blogs. Hence, talking about business is related to being at a high level in the organization. 

Three individuals (Actor 55, Actor 65, and Actor 94) were interpreted as either penetration testers, coders, or hackers by previous researchers. In our study, their dominant topic was the Technical topic, which relates to coding, testing, and hacking. Our results are thus consistent with previous research.   

Five actors were interpreted as managers with various specializations (see Table~\ref{tab:existing_analysis}) in previous analyses. In our analysis, the dominant topic of these actors was Internal Tasking and Management. This result is also consistent as it shows how managers, regardless of their specialization, are involved in internal and management tasks. 

On the other hand, three actors (Actor 85/Actor 132, and Actor 11) were interpreted as managers of technical teams in previous external analyses while, in our analysis, Malware is their dominant topic. These managers may thus have been more the type of technical/hands-on type of managers. Note that Actor 132 and Actor 85 are a pair in this table because they were referred to as being the same actor with two different pseudonyms ~\cite{etal_leaks_2022}. 

Finally, two actors (Actor 23 and Actor 36) had as a dominant topic Customer-Service/Problem Solving. One was interpreted as a technical manager responsible for coders in other blogs. The other was interpreted as a manager/Chief operating officer. These two roles align with having a high prevalence of Customer Service/Problem-Solving topics. 

While our results align with those from external sources, there are also some discrepancies. For instance, as shown in Table~\ref{tab:existing_analysis}, a Conti Chief Operation Officer's (COO) focus appears to be primarily on customer service and problem-solving (Actor 36). However, this COO was also classified as a  ``manager" by another source, showing discrepancies in role assignments from external sources. This is because assigning roles to individuals based on their conversations might not be perfectly accurate: what discussion topic an individual engages in depends on the individual's role but also on the individual's interests, skills, and the context in which the discussion takes place. The algorithm, on the other hand, produced a summary of the topics (most frequently co-occurring words) of actors' discussions, regardless of their roles. Further research could combine both methods to provide a more comprehensive link between roles and discussion topics.

\begin{table*}[ht]
    \centering
    \caption{Roles of Well-Known Actors and their Topic Distribution}
    \label{tab:existing_analysis}
    \begin{tabular}{|p{2.5cm}|c|p{2cm}|c|c|p{2cm}|c|p{2cm}|}
    \hline
        Roles~\cite{etal_leaks_2022, noauthor_conti_2022,yaaracyberintcom_be_2022,noauthor_conti_nodate} & Actor & Dominant Topic & Business & Technical & Internal Tasking and Management & Malware & Customer Service and Problem Solving \\ \hline
        Big boss & Actor 112 & Business & 98.5\% & 0.3\%  & 0.4\% & 0.5\% & 0.3\% \\ \hline
        Effective head of office operations & Actor 118 & Business &99.4\%& 0.1\% & 0.1\% & 0.2\% & 0.1\% \\ \hline
        Tester/coder & Actor 55 & Technical &5.1\% & 53.3\% &  17.2\% & 7.3\% & 17.1\% \\ \hline
        Penetration testing/hacking team leader & Actor 65 & Technical &1.2\% & 66.6\% & 1.0\% & 29.4\% & 1.7\% \\ \hline
        Coder & Actor 94 & Technical  & 7.2\% & 88.3\% & 1.1\% & 2.0\% & 1.3\% \\ \hline
        Emotet manager & Actor 132/Actor 85 & Internal Tasking and Management  & 8.4\% & 12.7\% & 44.8\% & 14.2\% & 19.9\% \\ \hline
        Reverse engineers manager & Actor 101 & Internal Tasking and Management & 24.8\% & 4.7\% & 57.3\% & 5.0\% & 8.1\% \\ \hline
        Senior manager & Actor 95 & Internal Tasking and Management & 20.2\% & 6.1\% & 66.6\% & 3.6\% & 3.4\% \\ \hline
        Team leader ransom operation & Actor 124 & Internal Tasking and Management& 9.1\%  & 2.2\% & 77.9\% & 3.6\% & 7.2\% \\ \hline
        Manager of general questions. Takes part in the HR process. & Actor 76 & Internal Tasking and Management& 15.8\%  & 0.5\% & 82.6\% & 0.8\% & 0.4\% \\ \hline
        Emotet manager & Actor 85/Actor 132 & Malware& 1.8\% & 9.3\% & 2.0\% & 84.9\% & 2.0\% \\ \hline
        Technical lead/manager of teams of crypters and testers & Actor 11 & Malware& 0.2\% & 0.2\% & 0.1\% & 99.4\% & 0.1\% \\ \hline
        Technical manager responsible for coders and their products, curating loaders and bots development within multiple coders teams & Actor 23 & Customer service and Problem-solving	& 10.7\%&	10.4\%&	22.0\%&	15.3\%&	41.6\%\\ \hline
         Manager/ COO &	 Actor 36 &	Customer service and Problem-solving &	13.2\%&	1.7\%&	1.3\%&	4.2\%&	79.6\% \\ \hline

\end{tabular}


Table~\ref{tab:existing_analysis} presents a comparative analysis of well-known actors' dominant topic found in our machine learning model and the roles reported in existing qualitative analysis by CheckPoint Research~\cite{etal_leaks_2022}, KrebsonSecurity ~\cite{noauthor_conti_2022}, Cyberint~\cite{yaaracyberintcom_be_2022}, and Trellix~\cite{noauthor_conti_nodate}. This comparison does not aim to establish the absolute truth regarding the actors' roles. It aims to assess the coherence of the results of this study, given other qualitative studies on the topic. It also shows the level of agreement and potential discrepancies between automated and human analysis methods.
\end{table*}

\section{Discussion}
\label{sec:discussion}

The results obtained are in line with large cybercrime organizations being organized similarly to firms~\cite{lusthaus_industry_2018}. This is highlighted by the three discussion points below: 1) the importance of non-tech talks, 2) culprit of specialization, yet diverse discussions, 3) higher-ups are business focus. The study results also corroborate key findings highlighted in previous qualitative research on the Conti RaaS operator~\cite{etal_leaks_2022,noauthor_conti_nodate-1,noauthor_conti_nodate,noauthor_conti_2022,kovacs_conti_2022}.

However, note that the Conti RaaS operator is one of the biggest RaaS operators and thus, this finding may be, in fact, an outlier. Whether a RaaS operator become organized as such probably depends on its size and scope as well as its success. Where members of a RaaS operator are located may also have an impact on its structure as places where the risks of arrests are low may facilitate the development of structured criminal organizations~\cite{lusthaus_industry_2018, lusthaus_offline_2021}. Further research should investigate other cybercrime organizations to see what influences their structure.

\textbf{The Importance of Non-Tech Talks}
The results of the study illustrate that a large proportion of discussions are non-technical and such discussion topics span across almost all Conti members. Non-tech talks encompass the Business and the Internal tasking/Management topics while focused tech talks encompassed the Malware and the Technical topics. The fifth topic, Customer Service/Problem Solving, included both. Merging non-tech talks and tech talks shows that, on average, 44.2\% (std=21.9) of actors' corpus involved non-tech talks, while 31.8\% (std=23.0) involved tech talks. The Customer Service/Problem Solving topic formed, on average, 24\% (std=17.6) of actors' corpus. These results show that Conti's daily operations required a lot of organization beyond writing malicious code to compromise networks. 

\textbf{Culprit of Specialization, Yet Diverse Discussions}
The results of the study also highlight that only a few members have a corpus that represented mainly a single topic. On the other hand, most actors in the dataset were diverse in their discussion topics: they mixed both Customer Service/Problem Solving with Internal tasking/Management as well as Business, Technical talks and Malware discussions. Hence, Conti's staff needed to work across multiple fields and have expertise in various areas, as highlighted in~\cite{etal_leaks_2022}. Moreover, as shown in Table~\ref{tab:existing_analysis}, some of Conti managers discussed about Customer Service/Problem Solving while others were more specialized, discussing more about Technical or Malware topics. Hence, some managers no longer talked as much about technical subjects, focusing instead on managing their team and dealing with customers. These different management roles were also noted in ~\cite{etal_leaks_2022}. Hence, although such RaaS operator represents the culprit of specialization in the cybercrime industry~\cite{salvi_raas_2019, meland_ransomware-as--service_2020, maurya_ransomware_2018, alwashali_survey_2021}, the bulk of its members appeared to have non-tech and diverse discussions, such discussions are likely required to coordinate the economic activities of a large criminal organization.

\textbf{Higher-ups are Business Focus}
According to ~\cite{etal_leaks_2022}, Conti higher-ups were always trying to find new ways to expand the firm's operation and produce more profit. Some of them even followed corporate tradition and held yearly performance review, talking about employees’ efficiency and deliberating on the employee of the month. Two actors' discussions revolved almost solely on Business topic : Actor 112 and Actor 118. As shown in Table ~\ref{tab:existing_analysis} Actor 112 and Actor 118 were identified as "Big Boss" and "Effective head of office operations" and both of their discussions revolved at 99\% around the Business topic. This finding supports the claim that Conti was indeed an organized firm with leaders constantly seeking fresh approaches to grow the company's activities and generate greater profits, as pointed out in~\cite{etal_leaks_2022}.

\section{Study Limitations and Future Research}
\label{sec:limits}
A first limitation of this study lies in the dataset as only the Jabber chat logs were used while the whole leak included also the rocket chat logs. To build on this limit, further studies could use the rocket chat logs or combine them with the jabber ones to investigate if the findings of this study hold with this additional corpus. Moreover, the original messages were written in Russian, and consequently, it is likely that the translations carried out was limited because of the use of Russian slang or abbreviations. Part of the meaning or nuance of a sentence may have been altered or lost through translation. Interpretation and reuse of results must take this limit into account. Consequently, it would also be interesting to carry out this research using the original chat logs in Russian. The use of the original chat logs would preserve all the meaning present in the data and could provide additional material and nuances the results.

Another limitation lies in the interpretation of the results. This study did not consider the size of the corpus, the timeline of the chats, nor the "member status" of the individuals. First, the corpus size may influence the topic distribution as individuals who discuss more may be more inclined to have generalist talks. Further studies should investigate how topic distribution influence the types of discussions in which individuals engage. Second, actors' experience in the organization was also not considered, limiting the interpretation of the results. For example, new individuals who have just arrived in the organization may have been more involved in specific discussion topics, such as human resources, due to their newcomer status. A more qualitative research focusing on the timeline of each actor could dive deeper into the data and analyze the changes of an actor over time. Third, this research does not consider the official status of the studied actors. Some actors are official members whereas others could be affiliates or even customers. Further studies should look whether topic distributions vary when considering the status of the actors studied.

A final limitation lies in the use of LDA models since such a model cannot capture contextual information, it only considers the frequency of words in a corpus. To overcome this, the conversations of actors having a specific topic as a dominant one were read and interpreted, thus providing contextual information around that topic. Further studies could deepen this analysis by conducting a qualitative thematic analysis of the conversations and comparing the results with this study.  

\section{Conclusion}
\label{sec:conclusion}
Leveraging the Conti chat leaks, this study uses machine learning algorithms to uncover insights on the organization of the Conti RaaS operator. The study shows that the discussions of the large RaaS operator Conti revolved around five topics: 1) Business, 2) Technical, 3) Internal tasking/Management, 4) Malware, and 5) Customer service/Problem Solving. Moreover, the topic distribution illustrates that only a few actors had specialized discussions in one topic, while the rest were all-rounders. The results corroborate that large cybercrime organizations are organized similarly to firms~\cite{lusthaus_industry_2018}. This is highlighted due to the importance of non-tech talks in the chats, the diverse discussion topics (although the organization represents the culprit of specialization), the varied management styles of actors, and how higher-ups, and specifically the two bosses, were business-focused in their discussions. Finally, this study illustrates how to automatically extract actionable information on the organization of a sophisticated cybercrime organization.

\section*{Acknowledgments}
This research was funded by the Human-Centric Cybersecurity Partnership (HC2P). We thank members of the Stratosphere Laboratory and the EconCrime Lab for reviewing previous versions of the article. We also thank Maxime Fuchs for his input during the brainstorm process.

\bibliographystyle{plain}
\bibliography{bibliography}

\begin{thebibliography}{10}

\bibitem{noauthor_conti_nodate-1}
Conti {Leaks}: {Examining} the {Panama} {Papers} of {Ransomware} {\textbar}
  {Trellix}.

\bibitem{noauthor_conti_nodate}
Conti ({Ryuk}) joins the ranks of ransomware gangs operating data leak sites.

\bibitem{noauthor_downloading_nodate}
Downloading {MALLET}.

\bibitem{noauthor_gensim_nodate}
gensim: {Python} framework for fast {Vector} {Space} {Modelling}.

\bibitem{noauthor_official_nodate}
Official {Alerts} \& {Statements} - {FBI} {\textbar} {CISA}.

\bibitem{noauthor_pyldavis_nodate}
{pyLDAvis} — {pyLDAvis} 2.1.2 documentation.

\bibitem{noauthor_vx-underground_nodate}
vx-underground - {Directory}.

\bibitem{noauthor_ransomware_2020}
Ransomware {Activity} {Targeting} the {Healthcare} and {Public} {Health}
  {Sector} {\textbar} {CISA}, November 2020.

\bibitem{noauthor_complete_2022}
The {COMPLETE} translation of leaked files related to {Conti} {Ransomware}
  group, October 2022.
\newblock original-date: 2022-03-04T16:13:44Z.

\bibitem{noauthor_conti_2022-1}
conti leaks sur {Twitter}, March 2022.

\bibitem{noauthor_conti_2022}
Conti {Ransomware} {Group} {Diaries}, {Part} {II}: {The} {Office} – {Krebs}
  on {Security}, March 2022.

\bibitem{alwashali_survey_2021}
Ali Ahmed Mohammed~Ali Alwashali, Nor Azlina~Abd Rahman, and Noris Ismail.
\newblock A {Survey} of {Ransomware} as a {Service} ({RaaS}) and {Methods} to
  {Mitigate} the {Attack}.
\newblock In {\em 2021 14th {International} {Conference} on {Developments} in
  {eSystems} {Engineering} ({DeSE})}, pages 92--96, December 2021.
\newblock ISSN: 2161-1351.

\bibitem{alzahrani_analysis_2022}
Saleh Alzahrani, Yang Xiao, and Wei Sun.
\newblock An {Analysis} of {Conti} {Ransomware} {Leaked} {Source} {Codes}.
\newblock {\em IEEE Access}, 10:100178--100193, 2022.
\newblock Conference Name: IEEE Access.

\bibitem{blei_latent_nodate}
David~M Blei.
\newblock Latent {Dirichlet} {Allocation}.

\bibitem{brewer_ransomware_2016}
Ross Brewer.
\newblock Ransomware attacks: detection, prevention and cure.
\newblock {\em Network Security}, 2016(9):5--9, September 2016.

\bibitem{chainalysis_ransomware_2023}
Team Chainalysis.
\newblock Ransomware {Revenue} {Down} {As} {More} {Victims} {Refuse} to {Pay},
  January 2023.

\bibitem{chowdhary_natural_2020}
K.~R. Chowdhary.
\newblock Natural {Language} {Processing}.
\newblock In K.R. Chowdhary, editor, {\em Fundamentals of {Artificial}
  {Intelligence}}, pages 603--649. Springer India, New Delhi, 2020.

\bibitem{etal_leaks_2022}
etal.
\newblock Leaks of {Conti} {Ransomware} {Group} {Paint} {Picture} of a
  {Surprisingly} {Normal} {Tech} {Start}-{Up}... {Sort} {Of}, March 2022.

\bibitem{gazet_comparative_2010}
Alexandre Gazet.
\newblock Comparative analysis of various ransomware virii.
\newblock {\em Journal in Computer Virology}, 6(1):77--90, February 2010.

\bibitem{henning_machine_nodate}
Andrew~Stephen Henning.
\newblock Machine {Learning} {And} {Natural} {Language} {Methods} {For}
  {Detecting} {Psychopathy} {In} {Textual} {Data}.

\bibitem{huang_systematically_2018}
Keman Huang, Michael Siegel, and Stuart Madnick.
\newblock Systematically {Understanding} the {Cyber} {Attack} {Business}: {A}
  {Survey}.
\newblock {\em ACM Computing Surveys}, 51(4):70:1--70:36, July 2018.

\bibitem{hyslip_cybercrime-as--service_2020}
Thomas~S. Hyslip.
\newblock Cybercrime-as-a-{Service} {Operations}.
\newblock In Thomas~J. Holt and Adam~M. Bossler, editors, {\em The {Palgrave}
  {Handbook} of {International} {Cybercrime} and {Cyberdeviance}}, pages
  815--846. Springer International Publishing, Cham, 2020.

\bibitem{kalaimannan_influences_2017}
Ezhil Kalaimannan, Sharon~K. John, Theresa DuBose, and Anthony Pinto.
\newblock Influences on ransomware’s evolution and predictions for the future
  challenges.
\newblock {\em Journal of Cyber Security Technology}, 1(1):23--31, January
  2017.
\newblock Publisher: Taylor \& Francis \_eprint:
  https://doi.org/10.1080/23742917.2016.1252191.

\bibitem{kamil_rise_2022}
Samar Kamil, Huda Sheikh~Abdullah Siti~Norul, Ahmad Firdaus, and Opeyemi~Lateef
  Usman.
\newblock The {Rise} of {Ransomware}: {A} {Review} of {Attacks}, {Detection}
  {Techniques}, and {Future} {Challenges}.
\newblock In {\em 2022 {International} {Conference} on {Business} {Analytics}
  for {Technology} and {Security} ({ICBATS})}, pages 1--7, February 2022.

\bibitem{kharraz_cutting_2015}
Amin Kharraz, William Robertson, Davide Balzarotti, Leyla Bilge, and Engin
  Kirda.
\newblock Cutting the {Gordian} {Knot}: {A} {Look} {Under} the {Hood} of
  {Ransomware} {Attacks}.
\newblock In Magnus Almgren, Vincenzo Gulisano, and Federico Maggi, editors,
  {\em Detection of {Intrusions} and {Malware}, and {Vulnerability}
  {Assessment}}, Lecture {Notes} in {Computer} {Science}, pages 3--24, Cham,
  2015. Springer International Publishing.

\bibitem{kirda_unveil_2017}
Engin Kirda.
\newblock {UNVEIL}: {A} large-scale, automated approach to detecting ransomware
  (keynote).
\newblock In {\em 2017 {IEEE} 24th {International} {Conference} on {Software}
  {Analysis}, {Evolution} and {Reengineering} ({SANER})}, pages 1--1, February
  2017.

\bibitem{kok_ransomware_2019}
S.~Kok, A.~Abdullah, Noor~Zaman Jhanjhi, and Mahadevan Supramaniam.
\newblock Ransomware , {Threat} and {Detection} {Techniques} : {A} {Review}.
\newblock 2019.

\bibitem{kolini_clustering_nodate}
Farzan Kolini and Lech Janczewski.
\newblock Clustering and {Topic} {Modelling}: {A} {New} {Approach} for
  {Analysis} of {National} {Cyber} security {Strategies}.
\newblock 2017.

\bibitem{kovacs_conti_2022}
Eduard Kovacs.
\newblock Conti {Ransomware} {Operation} {Shut} {Down} {After} {Brand}
  {Becomes} {Toxic}, May 2022.

\bibitem{lee_ransomware_2018}
Kyungroul Lee, Kangbin Yim, and Jung~Taek Seo.
\newblock Ransomware prevention technique using key backup.
\newblock {\em Concurrency and Computation: Practice and Experience},
  30(3):e4337, 2018.

\bibitem{leukfeldt_cybercrime_2014}
E.~R. Leukfeldt.
\newblock Cybercrime and social ties: {Phishing} in {Amsterdam}.
\newblock {\em Trends in Organized Crime}, November 2014.

\bibitem{leukfeldt_examining_2020}
E.~R. Leukfeldt and Thomas~J. Holt.
\newblock Examining the {Social} {Organization} {Practices} of {Cybercriminals}
  in the {Netherlands} {Online} and {Offline}.
\newblock {\em International Journal of Offender Therapy and Comparative
  Criminology}, 64(5):522--538, April 2020.
\newblock Publisher: SAGE Publications Inc.

\bibitem{leukfeldt_criminal_2019}
E.~Rutger Leukfeldt, Edward~R. Kleemans, Edwin~W. Kruisbergen, and Robert~A.
  Roks.
\newblock Criminal networks in a digitised world: on the nexus of borderless
  opportunities and local embeddedness.
\newblock {\em Trends in Organized Crime}, 22(3):324--345, September 2019.

\bibitem{leukfeldt_cybercriminal_2016}
E.~Rutger Leukfeldt, Edward~R. Kleemans, and Wouter~P. Stol.
\newblock Cybercriminal {Networks}, {Social} {Ties} and {Online} {Forums}:
  {Social} {Ties} {Versus} {Digital} {Ties} within {Phishing} and {Malware}
  {Networks}.
\newblock {\em British Journal of Criminology}, page azw009, February 2016.

\bibitem{leukfeldt_origin_2017}
E.~Rutger Leukfeldt, Edward~R. Kleemans, and Wouter~P. Stol.
\newblock Origin, growth and criminal capabilities of cybercriminal networks.
  {An} international empirical analysis.
\newblock {\em Crime, Law and Social Change}, 67(1):39--53, February 2017.

\bibitem{leukfeldt_typology_2017}
E.~Rutger Leukfeldt, Edward~R. Kleemans, and Wouter~P. Stol.
\newblock A typology of cybercriminal networks: from low-tech all-rounders to
  high-tech specialists.
\newblock {\em Crime, Law and Social Change}, 67(1):21--37, February 2017.

\bibitem{leukfeldt_organised_2017}
E.~Rutger Leukfeldt, Anita Lavorgna, and Edward~R. Kleemans.
\newblock Organised {Cybercrime} or {Cybercrime} that is {Organised}? {An}
  {Assessment} of the {Conceptualisation} of {Financial} {Cybercrime} as
  {Organised} {Crime}.
\newblock {\em European Journal on Criminal Policy and Research},
  23(3):287--300, September 2017.

\bibitem{leukfeldt_use_2017}
Rutger Leukfeldt, Edward Kleemans, and Wouter Stol.
\newblock The {Use} of {Online} {Crime} {Markets} by {Cybercriminal}
  {Networks}: {A} {View} {From} {Within}.
\newblock {\em American Behavioral Scientist}, 61(11):1387--1402, October 2017.

\bibitem{lusthaus_industry_2018}
Jonathan Lusthaus.
\newblock {\em Industry of {Anonymity}: {Inside} the {Business} of
  {Cybercrime}}.
\newblock Harvard University Press, \$ \{nombre\}er édition edition, October
  2018.

\bibitem{lusthaus_offline_2021}
Jonathan Lusthaus and Federico Varese.
\newblock Offline and {Local}: {The} {Hidden} {Face} of {Cybercrime}.
\newblock {\em Policing: A Journal of Policy and Practice}, 15(1):4--14, May
  2021.

\bibitem{manky_cybercrime_2013}
Derek Manky.
\newblock Cybercrime as a service: a very modern business.
\newblock {\em Computer Fraud \& Security}, 2013(6):9--13, June 2013.

\bibitem{maurya_ransomware_2018}
A.K. Maurya, Neeraj Kumar, Alka Agrawal, and Prof.~Raees Khan.
\newblock Ransomware {Evolution}, {Target} and {Safety} {Measures}.
\newblock {\em International Journal of Computer Sciences and Engineering},
  6:80--85, January 2018.

\bibitem{meland_ransomware-as--service_2020}
Per~Håkon Meland, Yara Fareed~Fahmy Bayoumy, and Guttorm Sindre.
\newblock The {Ransomware}-as-a-{Service} economy within the darknet.
\newblock {\em Computers \& Security}, 92:101762, May 2020.

\bibitem{ofer_language_2021}
Dan Ofer, Nadav Brandes, and Michal Linial.
\newblock The language of proteins: {NLP}, machine learning \& protein
  sequences.
\newblock {\em Computational and Structural Biotechnology Journal},
  19:1750--1758, January 2021.

\bibitem{okane_evolution_2018}
Philip O'Kane, Sakir Sezer, and Domhnall Carlin.
\newblock Evolution of ransomware.
\newblock {\em IET Networks}, 7(5):321--327, 2018.

\bibitem{oosthoek_tale_2022}
Kris Oosthoek, Jack Cable, and Georgios Smaragdakis.
\newblock A {Tale} of {Two} {Markets}: {Investigating} the {Ransomware}
  {Payments} {Economy}, May 2022.
\newblock arXiv:2205.05028 [cs].

\bibitem{peattie1995approaching}
Noel Peattie.
\newblock Approaching zero: The extraordinary underworld of hackers, phreakers,
  virus writers, and keyboard criminals.
\newblock {\em Journal of Information Ethics}, 4(2):79, 1995.

\bibitem{raina_natural_2022}
Vineet Raina and Srinath Krishnamurthy.
\newblock Natural {Language} {Processing}.
\newblock In Vineet Raina and Srinath Krishnamurthy, editors, {\em Building an
  {Effective} {Data} {Science} {Practice}: {A} {Framework} to {Bootstrap} and
  {Manage} a {Successful} {Data} {Science} {Practice}}, pages 63--73. Apress,
  Berkeley, CA, 2022.

\bibitem{reuter_disorganized_1983}
Peter Reuter.
\newblock {\em Disorganized {Crime}: {Illegal} {Markets} and the {Mafia}}.
\newblock Organization {Studies} series. MIT Press, Cambridge, MA, USA, March
  1983.

\bibitem{richardson_ransomware_2017}
Ronny Richardson and M.~North.
\newblock Ransomware: {Evolution}, {Mitigation} and {Prevention}.
\newblock {\em International Management Review}, 2017.

\bibitem{ryan_ransomware_2021}
Matthew Ryan.
\newblock {\em Ransomware {Revolution}: {The} {Rise} of a {Prodigious} {Cyber}
  {Threat}}, volume~85 of {\em Advances in {Information} {Security}}.
\newblock Springer International Publishing, Cham, 2021.

\bibitem{salvi_raas_2019}
Harshada~Umesh Salvi.
\newblock {RAAS}: {Ransomware}-as-a-{Service}.
\newblock {\em International Journal of Computer Sciences and Engineering},
  7(6):586--590, June 2019.

\bibitem{scaife_cryptolock_2016}
Nolen Scaife, Henry Carter, Patrick Traynor, and Kevin R.~B. Butler.
\newblock {CryptoLock} (and {Drop} {It}): {Stopping} {Ransomware} {Attacks} on
  {User} {Data}.
\newblock In {\em 2016 {IEEE} 36th {International} {Conference} on
  {Distributed} {Computing} {Systems} ({ICDCS})}, pages 303--312, June 2016.
\newblock ISSN: 1063-6927.

\bibitem{song_effective_2016}
Sanggeun Song, Bongjoon Kim, and Sangjun Lee.
\newblock The {Effective} {Ransomware} {Prevention} {Technique} {Using}
  {Process} {Monitoring} on {Android} {Platform}.
\newblock {\em Mobile Information Systems}, 2016:e2946735, May 2016.
\newblock Publisher: Hindawi.

\bibitem{theparmak_conti-leaks-englished_2023}
TheParmak.
\newblock conti-leaks-englished, February 2023.

\bibitem{umar_analysis_2021}
Rusydi Umar, Imam Riadi, and Ridho~Surya Kusuma.
\newblock Analysis of {Conti} {Ransomware} {Attack} on {Computer} {Network}
  with {Live} {Forensic} {Method}.
\newblock {\em IJID (International Journal on Informatics for Development)},
  10(1):53--61, June 2021.
\newblock Number: 1.

\bibitem{vashisht_integrating_2020}
Vipul Vashisht and Pankaj Dharia.
\newblock Integrating {Chatbot} {Application} with {Qlik} {Sense} {Business}
  {Intelligence} ({BI}) {Tool} {Using} {Natural} {Language} {Processing}
  ({NLP}).
\newblock In Devendra~Kumar Sharma, Valentina~Emilia Balas, Le~Hoang Son, Rohit
  Sharma, and Korhan Cengiz, editors, {\em Micro-{Electronics} and
  {Telecommunication} {Engineering}}, Lecture {Notes} in {Networks} and
  {Systems}, pages 683--692, Singapore, 2020. Springer.

\bibitem{wu_ranking_2011}
Yonghui Wu, Mei Liu, W.~Jim Zheng, Zhongming Zhao, and Hua Xu.
\newblock Ranking gene-drug relationships in biomedical literature using latent
  dirichlet allocation.
\newblock In {\em Biocomputing 2012}, pages 422--433. WORLD SCIENTIFIC,
  November 2011.

\bibitem{yaaracyberintcom_be_2022}
yaara@cyberint.com.
\newblock To {Be} {CONTInued}? {Conti} {Ransomware} {Heavy} {Leaks}, March
  2022.

\bibitem{zhou_comparative_2019}
Yanfen Zhou and Jin-Cheon Na.
\newblock A comparative analysis of {Twitter} users who {Tweeted} on psychology
  and political science journal articles.
\newblock {\em Online Information Review}, 43(7):1188--1208, January 2019.
\newblock Publisher: Emerald Publishing Limited.

\bibitem{zong_application_2018}
Zhaorong Zong and Changchun Hong.
\newblock On {Application} of {Natural} {Language} {Processing} in {Machine}
  {Translation}.
\newblock In {\em 2018 3rd {International} {Conference} on {Mechanical},
  {Control} and {Computer} {Engineering} ({ICMCCE})}, pages 506--510, September
  2018.

\end{thebibliography}

\end{document}